\documentclass[reprint,onecolumn,superscriptaddress,amsmath,amssymb,aps,pra,nofootinbib]{revtex4-2}

\makeatletter
\gdef\@ptsize{0}% 10pt documents
\let\@currsize\normalsize
\makeatother
\usepackage{setspace}
\doublespace

\usepackage{mathtools}
\usepackage{graphicx}

\usepackage{amssymb}

\begin{document}

\title
{Is there any reasonable dynamics for quantum subsystems\\
 within the unitary-only quantum theory?}

\author{Miroljub \surname{Dugi\'c}}\email{mdugic18@sbb.rs}
\affiliation{University of Kragujevac, Faculty of Science, Radoja
Domanovi\' ca 12, 34000 Kragujevac, Serbia}
\author{Jasmina \surname{Jekni\' c-Dugi\' c}}\email{jjeknic@pmf.ni.ac.rs}
\affiliation{University of Ni\v s, Faculty of Science and Mathematics, Vi\v
segradska 33, 18000 Ni\v s, Serbia}
\author{Momir \surname{Arsenijevi\'c}}\email{fajnman@gmail.com}
\affiliation{University of Kragujevac, Faculty of Science, Radoja
Domanovi\' ca 12, 34000 Kragujevac, Serbia}

\date{\today}

\begin{abstract} The foundational studies of the standard, unitary-only quantum theory revolve around the kinematical aspects of quantum entanglement and the improper quantum mixtures. In this paper, we introduce and argue for the foundational character of the question of  dynamics of  quantum subsystems (open quantum systems). In this context, for some typical and physically relevant Markovian processes, we technically prove non-existence of trajectories in the Hilbert state space of the open system. As a kind of no-go theorem for the unitary-only quantum theory, this finding suggests that the mixed quantum states may be joined to the individual (single) quantum subsystem dynamically described by the corresponding master equation. Then the problem of interpretation of improper mixtures dissolves while  description of quantum measurement boils down to the problem of reduction of the mixed to the pure states--i.e. to the problem of actualization of  definite values of certain observables of the single open systems, thus tackling the mathematical problem of interpreting probability for the single trials of an experiment. This kind of indeterminism may be the furthest we can go within the dynamical approach to quantum subsystems. As an alternative appears the possibility that the idea of dynamics of quantum subsystems may not be viable in the context of the unitary-only theory. As a direct consequence of our main finding appears the a priori impossibility to define ''quantum history'' in the Hilbert state space for the considered Markovian models.
\end{abstract}

\maketitle

\section{Introduction}

It is a tacit assumption of the standard (unitary-only) \cite{foot,kastner} quantum mechanical theory: for a statistical ensemble in a pure state $\vert\psi\rangle$, every single, i.e. every individual, element of the ensemble is in the same pure state $\vert\psi\rangle$. Then, if every individual element of the ensemble is subject of the same dynamics, every individual element of the ensemble is in the same state in every instant of time. Bearing in mind that phenomenology demonstrates the opposite, e.g., for the object of measurement, it seems unavoidable to conclude that the different individual
objects of measurement must have gone different dynamics. Then the question, equally targeting the individual as well as the statistical ensembles of such systems, arises:

\noindent (Q) What might be the dynamical law(s) for a quantum subsystem?

Within the standard, unitary-only quantum theory, a quantum subsystem is described by the so-called ''improper mixture'', which cannot be recognized as the true quantum state for either individual systems, or an ensemble of such systems \cite{Despa}. Then it is hard to imagine how quantum dynamics could be defined for quantum subsystems, both on the level of the single (individual) systems as well as of a statistical ensemble of such systems.

The problem becomes even more severe for the open systems initially correlated with their environments, in which case the initial state of the open system is formally defined by ''improper mixture''. To this end, some researchers argued for ill-defined dynamics \cite{pechukas}, while some others \cite{alicki,shaji} argued for the possibly non-completely-positive or even non-linear dynamics of the open systems initially correlated with their environments. Within some recent proposals, it can be found the argument that theoretical considerations may be subject of insufficient tools to properly perceive the problem, while the experimental work is endowed by measurement-based acquisition of data and therefore essentially free of the controversy \cite{milz}. The told insufficiency may be due to discarding the very concept of  state of an initially entangled subsystem \cite{milz} or avoiding the issue of the subsystem dynamics \cite{gisinnovo}.
It may be a matter of taste how far have we gone in those contexts from the problem at issue--the problem of interpretation of ''improper mixtures''.

The problem at issue is implicit to the task of solving, e.g., Markovian master equations for open systems. A recent proposal termed ''ensemble rank truncation'' (ERT) employs an approximation of the process in the Kraus form and in each step of calculation provides a set of pure states (''wave functions'') that sufficiently well approximates the exact solution of the system's density matrix (statistical operator) \cite{McCaul}. Nevertheless, the absence of {\it dynamical} links between the pure states in the consecutive calculation steps is symptomatic for the problem at issue. That is, discarding certain pure states due to the truncation does not allow for detecting the pure state trajectories, $\vert\Psi(t)\rangle$, in the Hilbert state space of the system and thus the possibility to introduce the open-system's dynamics. Interestingly, the ERT method may be regarding a computational alternative to the methods employing an external source of classical noise and therefore abandoning the framework of the unitary-only theory \cite{heger}.

Before abandoning the unitary-only (theoretical) framework, the following observation may be in order. In quantum measurement as well as in the general Open systems theory \cite{breuer, rivas}, quantum state of every individual element of the initial ensemble in pure state is typically assumed to be well defined. It is only the {\it interaction} of an open system (e.g. of an object of measurement), denoted $S$, with some other quantum system(s), denoted $E$, that introduces an improper mixture for the $S$ system and therefore the above posed question. Having said that, the following question naturally arises: may it be the case that the quantum puzzle imposed by ''improper mixtures'' is {\it dynamical} at the root, in the sense that the lack of knowledge of the subsystem’s dynamics may be the origin of the problem, while the standard kinematical considerations tackling the instantaneous improper mixtures for the $S$ system may be an effective description that would follow from a deeper dynamical description of the open system? More formally: instead of the quantum states [as typical for the kinematical approach], in the focus of our interest is placed the subsystem’s {\it dynamical map}, denoted $\mathcal{G}(t,0)$.  In contrast to the approach proposed in \cite{milz}, we assume that the initial state of an ensemble of the open systems is well defined.

In this paper, within the unitary-only quantum theory, we technically show that, for some typical Markovian open quantum systems, it is {\it in principle} not {\it possible to define trajectory in the Hilbert state space of the open system}. More technically: while the open system’s statistical operator (density matrix) evolution in time is described by a Markovian master equation, there is not even a single initial pure state that could sustain its purity in the course of the system’s dynamics. To the best of our knowledge there is not a general result that could {\it a priori} reject this pure pure-state dynamics (PPSD) of the open systems (Markovian or not)—see Section II for more information. In this paper we use a general criterion for the PPSD of Markovian systems \cite{sandu} and still do not provide the general result(s) that could apply to all Markovian processes—not even to mention the non-Markovian dynamics.  We want strongly to stress that the considered models regard some of the physically profound models of interest in the Open quantum systems theory \cite{breuer,rivas}, Quantum measurement and decoherence theory \cite{Despa,Giulini,Slosi} and applications (such as e.g. Quantum optics \cite{Gardiner,Scully}). That is, our conclusions do not stem from some artificial or pathological models.

	Per se, our finding is a kind of no-go theorem for the only-unitary quantum theory: a priori, there is not a single system (or a subensemble) initially in a pure state that could dynamically remain in a pure state for the considered models. The fact that we deal with the dynamical aspects of quantum subsystems (open systems) instead of the standard kinematical considerations of entangled bipartite systems \cite{Nilsen} makes our approach irreducible to the standard kinematical analysis of improper mixtures.

In the face of those statements, we formulate and briefly elaborate on a specific proposal for dealing with the question (Q) and the general description of the quantum measurement process, starting from the assumption that {\it even a single physical (sub)system may be described  by mixed states}. Then the problem of interpretation of improper mixtures dissolves, dynamics of single quantum open systems is formally well defined and the quantum measurement appears in a new guise. However, the price to be paid is that our proposal leads to the deeper problem of interpretation of probability for single systems (single elements of a statistical ensemble). An alternative might be that, within the unitary-only quantum theory, the concept of dynamics cannot be even introduced for quantum subsystems.

In Section II, we provide the main technical result of this paper. On this basis, we formulate our proposal in some detail in Section III. Section IV provides  comparison of our proposal with some prominent alternative theories/interpretations of the standard, unitary-only quantum theory. Section V is our conclusion.

\section{Trajectories in the Hilbert state space of an open quantum system}

	It is worth repeating: within the unitary-only quantum theory, the concept of improper mixtures for quantum subsystem—which is solely due to its entanglement with the external system $E$--is vague to the extent that it does not offer even a hint on the subsystem’s (open system’s) dynamics. Therefore the question (Q) posed in the previous section.
The so-called unital quantum processes, which preserve the maximally mixed state $\hat\rho=\hat I/d$ (where $d$ is dimension of the Hilbert space), are known to be the only kind of Markovian processes for which the so-called purity of a state $\hat\rho$, defined as $p:=tr(\hat\rho^2)$, satisfies \cite{lidar}:

\begin{equation}\label{eqn:1}
\dot{p} \equiv {d\over dt} tr(\dot{\hat{\rho}}^2) \le 0.
\end{equation}

\noindent That is, the unital processes are the only kind of Markovian dynamics (to be defined below) that does not increase purity of the quantum state in any instant $t$ of time. Equality in eq. (1) corresponds to the preservation of the initially pure state--that we are interested in. On the other hand, some non-unital Markovaian processes are known to have a pure state as the asymptotic state, $\hat\rho_{\infty}:=\lim_{t\to\infty}\hat\rho(t), \hat\rho_{\infty}^2=\hat\rho_{\infty}$. Therefore the pure pure-state dynamics (PPSD), i.e. the condition $\hat\rho^2(t)=\hat\rho(t)$ for the finite values of $t$, is not rejected yet on the general ground.

In this section we investigate the PPSD, i.e. existence of trajectories in the system's Hilbert state space, for some typical Markovian processes. It is worth stressing: nonexistence of trajectories in the Hilbert state space is qualitatively established \cite{Gisin1}  for certain models \cite{kent, benet} of deterministic nonlinear quantum dynamics and therefore cannot be used as an argument in our considerations.

In this section, we prove the following statement:

\smallskip

\noindent (S) At least for the considered models of Markovian open system dynamics, the pure pure-state dynamics providing a trajectory in the Hilbert state space of the open system is {\it not} possible.

\subsection{Two simple examples}

We start with two simple models of quantum decoherence since they can be analytically solved by providing a state of the system in explicit mathematical form.

Consider a single qubit (two-level) system subject of the quantum decoherence (quantum measurement of the $\hat Z$ Pauli operator) process as described by the master equation (in the interaction picture and neglecting the Lamb and the Stark shifts)\footnote{Sufficiency to work in the interaction picture will be emphasized in Section II.C.}:

\begin{equation}\label{eqn:2}
{d\hat\rho\over dt} = \gamma \left(\hat Z\hat \rho\hat Z-\hat\rho  \right).
\end{equation}

It is easy to obtain the solution to eq.(\ref{eqn:2}):

\begin{equation}\label{eqn:3}
\hat\rho(t) = \begin{pmatrix}
\rho_{11}(0) & \rho_{10}(0)e^{-2\gamma t} \\
\rho_{01}(0)e^{-2\gamma t} & 1-\rho_{11}(0)
\end{pmatrix}.
\end{equation}

\noindent for the initial state $\hat\rho(0) = \begin{pmatrix}
\rho_{11}(0) & \rho_{10}(0) \\
\rho_{01}(0) & 1-\rho_{11}(0)
\end{pmatrix}$, while $\hat Z\vert 1\rangle=\vert 1\rangle$ and $\hat Z\vert 0\rangle = -\vert 0\rangle$.

It is obvious from eq.(\ref{eqn:3}): for every $\rho_{11}(0)$ and $\rho_{10}(0)=0$, the state is stationary, i.e. it does not change with time--this is also the asymptotic limit, i.e. $\lim_{t\to\infty}\hat\rho(t) = \begin{pmatrix}
\rho_{11}(0) & 0 \\
0 & 1-\rho_{11}(0)
\end{pmatrix}$, as it can be expected: the "pointer basis states" are the eigenstates of the measured $\hat Z$ observable, which is also the ''pointer observable'' for the process. The point is that, assuming $\hat\rho(0)=\hat\rho^2(0)$, the state is mixed, i.e. $\hat\rho(t)\neq\hat\rho^2(t)$ for {\it any} finite instant of time $t$. That is, from eq.(\ref{eqn:3}) it is obvious that the initially pure state turns into a mixed state without purification for any finite $t$--there is not even a single trajectory in the Hilbert state space of the system.

Now let us consider a one-dimensional continuous variable (CV) system defined by a single Descartes degree of freedom, the position operator $\hat x$. Assume that the system is subject of the position decoherence (i.e. subject of the position measurement) as described by the master equation (in the interaction picture, neglecting the Lamb and the Stark shifts):

\begin{equation}\label{eqn:4}
{d\hat\rho\over dt} = \gamma\left(\hat x\hat\rho\hat x - (\hat x^2\hat\rho+\hat\rho\hat x^2)/2  \right).
\end{equation}

Then, in the position representation, $\rho(x,x')\equiv\langle x\vert\hat\rho\vert x'\rangle$, it is easy to obtain:

\begin{equation}\label{eqn:5}
{d\rho(x,x',t)\over dt} = -\gamma(x-x')^2\rho(x,x',t),
\end{equation}

\noindent thus providing unchange of the diagonal terms $\rho(x,x), \forall{x}$, and decay of the off-diagonal terms, $\rho(x,x',t)=\rho(x,x',0)e^{-\gamma t(x-x')^2}$. It  is probably obvious that eigenstates of $\hat x$, the unnormalizable $\vert x\rangle$ states, are exact "pointer basis" states and the observable $\hat x$ being the ''pointer observable'' for the process. The point to be emphasized is that, if the initial state is pure, i.e. $\rho(x,x',0)=\psi(x)\psi^{\ast}(x')$, the decay of the off-diagonal terms (like in the previous qubit example) implies $\hat\rho(t)\neq\hat\rho^2(t)$. That is, the initially pure state dynamically becomes mixed and does not purify for any finite instant of time $t$. In other words: there is not even a single trajectory in the Hilbert state space of the system for the considered process of the position measurement (decoherence).

\subsection{The general PPSD condition}

The models of Section II.A are special in that they admit analytical solutions to the master equations. In general, that may be a demanding task \cite{McCaul,braz1,braz2}. Fortunately, there is a general criterion for the pure pure-state dynamics (PPSD) for the Markovian processes \cite{sandu,isar}.

By Markovian processes, we assume the so-called CP-divisible processes\cite{rivas} that admit the following Lindblad form of the master equation (in the Schr\" odinger picture):

\begin{equation}\label{eqn:6}
{d\hat\rho(t)\over dt} \equiv \mathcal{L}[\hat\rho(t)] =  -{i\over \hbar} [\hat H,\hat\rho(t)] + \sum_i \gamma_i \left(\hat L_i \hat\rho(t) \hat L_i^{\dag} - {1\over 2}\{\hat L_i^{\dag} \hat L_i,\hat\rho(t)\}\right);
\end{equation}

\noindent in general, the damping factors $\gamma_i\ge 0$ and all the Lindblad operators ($\hat L_i$) may carry time dependence, while the time dependence of the Hermitian operator $\hat H$ may be allowed in the case of the week external field(s) \cite{rivas}.

 {\it If assumed to preserve purity} of the states in the course of the dynamics, equation (\ref{eqn:6}) takes the following form  \cite{sandu}:

\begin{equation}\label{eqn:7}
{d\over dt} (\vert\psi\rangle\langle\psi\vert) = -\imath (\mathcal{H}\vert\psi\rangle\langle\psi\vert - \vert\psi\rangle\langle\psi\vert \mathcal{H}^{\dag}),
\end{equation}

\noindent where:

\begin{equation}\label{eqn:8}
\mathcal{H} = \hat H + \imath \sum_i \gamma_i \left(
\langle \hat L_i^{\dag}\rangle \hat L_i - {1\over 2} \langle \hat L_i^{\dag}\hat L_i\rangle - {1\over 2} \hat L_i^{\dag} \hat L_i
\right),
\end{equation}

\noindent while $\langle\ast\rangle \equiv \langle\psi\vert\ast\vert\psi\rangle = tr(\vert\psi\rangle\langle\psi\vert\ast)$ and we assume $\hbar=1$.
A derivation of equations (\ref{eqn:7}) and (\ref{eqn:8}) for the semigroup maps can be found in \cite{sandu}; the main steps in the derivation, also applying for the general Markovian case, are presented in Appendix A.

Now, the PPSD process is defined by the condition $\hat\rho^2(t)=\hat\rho(t), \forall{t}$, where the time interval $[0,t]$ is arbitrary. Equivalently, the PPSD is defined by equality in eq.(1), i.e. by the condition $tr(\hat\rho d\hat\rho/dt)=0$. Then it can be shown that the necessary  condition for PPSD reads \cite{sandu}:

\begin{equation}\label{eqn:9}
\sum_i \gamma_i \langle\psi\vert \hat L_i\vert\psi\rangle\langle\psi\vert  \hat L_i^{\dag} \vert \psi\rangle =  \sum_i \gamma_i   \langle\psi\vert \hat L_i^{\dag}\hat L_i\vert\psi\rangle,
\end{equation}

\noindent where appear the Lindblad operators $\hat L_i$ from eq.(\ref{eqn:6}).

Existence of at least two states $\vert\psi(t)\rangle$ satisfying eq.(\ref{eqn:9}) is necessary for existence of at least two trajectories in the Hilbert state space, while those states are also required to satisfy  equation (\ref{eqn:7}). Without fulfillment of any of these conditions,  trajectories in the Hilbert space of the system do {\it not} exist\footnote{Approximate equalities in equations (\ref{eqn:7}) and (\ref{eqn:9}) introduce approximate purity of the states that fall out of the system's Hilbert state space and therefore are of no interest for our considerations.}.

\subsection{Analysis of some Markovian processes}

Consider the following two paradigmatic  models  of the homogeneous Markovian processes in the context of the condition eq.(\ref{eqn:9}). Due to the condition eq.(\ref{eqn:9}), we will regard only the dissipator term, which is for Markovian processes the same in the Schr\" odinger and the interaction picture, thus resorting to the latter.

\noindent --{\it Damping of a two-level system.} Choose the system's basis states denoted $\vert \pm\rangle$ and construct the observable $\hat\sigma_z \vert\pm\rangle=\pm\vert\pm\rangle$. Introduce also the nonhermitian operators $\hat\sigma_-=\vert -\rangle\langle +\vert =\hat\sigma_+^{\dag}$. Then consider the following homogeneous Markovian master equation (in the interaction picture) for such system \cite{breuer,rivas}:

\begin{equation}\label{eqn:10}
{d\hat\rho\over dt} = \gamma_1 \left(\hat\sigma_+ \hat\rho \hat\sigma_- -{1\over 2} \{\hat\sigma_-\hat\sigma_+,\hat\rho\}\right) + \gamma_2 \left(\hat\sigma_- \hat\rho \hat\sigma_+ -{1\over 2} \{\hat\sigma_+\hat\sigma_-,\hat\rho\}\right)
\end{equation}

\noindent where the curly brackets denote the anticommutator. Physically, for $\gamma_1=0$, we have the case of a single-qubit amplitude damping process \cite{Nilsen}, or the model of a two-level atomic system in contact with the environment on the absolute zero temperature \cite{breuer,rivas}. For the thermal bath of mutually uncoupled bosonic modes, the damping factors can read \cite{breuer, rivas}:

\begin{equation}\label{eqn:11}
\gamma_1=\gamma_{\circ}N, \gamma_2 = \gamma_{\circ}(N+1),
\end{equation}

\noindent where

\begin{equation}\label{eqn:12}
N=(e^{\beta\hbar\omega_{\circ}}-1)^{-1}\ge 0,
\end{equation}

\noindent is the mean number of quanta for the  mode of frequency $\omega_{\circ}$ and the bath on the inverse temperature $\beta$, while the real $\gamma_{\circ}>0$.

Let us separately consider the two cases: when $\gamma_1=0$, and when $\gamma_i\neq 0, i=1,2$.

\noindent {\it The case $\gamma_1=0$}. For this case of the absolute zero, $T=0$ (and $N=0$), the master equation (11) reduces to the second term, $\gamma_2=\gamma_0$, with the only one Lindblad operator, $\hat L=\hat\sigma_-$. Then eq.(\ref{eqn:9}) easily gives the pure-state dynamics condition to read:

\begin{equation}\label{eqn:13}
p_+(1-p_+)=p_+,
\end{equation}

\noindent where $p_+\equiv \vert\langle +\vert \psi\rangle\vert^2$. That is, eq.(14) reveals the only one possible solution: $p_+=0$ and therefore the only  one pure state allowed for every instant of time--the
ground state $\vert -\rangle$, which is easily shown to be a stationary state for the process that does not evolve in time.

\noindent {\it The case} $\gamma_i\neq 0,i=1,2$. Now consider the full master equation (\ref{eqn:10}) with both nonzero damping factors $\gamma_i$. In this case there are the two Lindblad operators, $\hat\sigma_-$ and $\hat\sigma_+$.
Due to Section II.A and nonexistence of even a single common eigenstate for $\hat\sigma_+$ and $\hat\sigma_-$,
 we know in advance that the pure-state condition eq.(6) cannot be satisfied. Nevertheless, for the sake of completeness as well as of demonstrating the condition (\ref{eqn:9}) for this model, we proceed as follows.

A straightforward application of eq.(\ref{eqn:9}) with the definitions eq.(12) and (13)  gives the pure-state dynamics condition:

\begin{equation}\label{eqn:14}
(2N+1)p_+^2 - 2Np_+ + N = 0,
\end{equation}

\noindent which leads to the solutions for the $p_+$:

\begin{equation}\label{eqn:15}
p_{+1,2} = {N\pm \sqrt{-N^2-N}\over2N+1}.
\end{equation}

Again, from eq.(16) follows the {\it only one} solution: $N=0$, which implies $p_+=0,\forall{t}$. Needless to say, $N=0$ is equivalent with $T=0$--which is analysed above--thus emphasizing the absence of even a single pure state satisfying eq.(\ref{eqn:9}) for the finite temperature.

\smallskip

\noindent --{\it The damped linear harmonic oscillator.} The one-dimensional (i.e. one-mode) harmonic oscillator damped by the bosonic heat bath is described by the widely used master equation (in the interaction picture) \cite{breuer,rivas}:

\begin{equation}\label{eqn:16}
{d\hat\rho\over dt} = \gamma_{\circ}(N+1) \left(
\hat a\hat\rho \hat a^{\dag} - {1\over 2} \{\hat a^{\dag}\hat a,\hat\rho\}
\right) + \gamma_{\circ}N \left(
\hat a^{\dag}\hat\rho \hat a- {1\over 2}\{\hat a \hat a^{\dag},\hat\rho\}
\right),
\end{equation}

\noindent where appear the standard bosonic operators, $[\hat a, \hat a^{\dag}]=\hat I$, with the $N$ defined by eq.(13), while the real $\gamma_{\circ}>0$.

\noindent {\it The case $T=0$}. Then $N=0$ and hence the master equation reduces to the first term with only one Lindblad operator, $\hat L=\hat a$. For this case the condition eq.(\ref{eqn:9}) reads:

\begin{equation}\label{eqn:17}
\langle\psi\vert \hat a\vert\psi\rangle \langle\psi\vert \hat a^{\dag}\vert\psi\rangle = \langle\psi\vert \hat a^{\dag} \hat a\vert\psi\rangle.
\end{equation}

It is probably obvious that every ''coherent state'' $\vert \alpha\rangle$
satisfying the eigen-problem $\hat a\vert\alpha\rangle = \alpha\vert\alpha\rangle$ satisfies the condition eq.(18) for the pure-state dynamics. Therefore it may seem possible to have trajectories of the ''coherent states'' $\vert\alpha(t)\rangle$ continuously transforming into each other.

However, placing a coherent state $\vert\alpha(t)\rangle$ into eq.(\ref{eqn:7}) leads to non-satisfiability of eq.(\ref{eqn:7}). Possibly the simplest way to show this is to employ the Sudarshan-Glauber $P$-representation, $\hat\rho(t) = \int d^2\alpha P(\alpha,\alpha^{\ast},t) \vert\alpha\rangle\langle\alpha\vert$, with the quasi-probability distribution $P(\alpha,\alpha^{\ast},t)$ satisfying $\int d^2\alpha P(\alpha,\alpha^{\ast},t)=1$. Equation (\ref{eqn:7}) assumes the constraint of purity of the state, $P(\alpha,\alpha^{\ast},t)=\delta^{(2)}(\alpha-\alpha(t)), \forall{t}$; $\delta^{(2)}(\ast)$ is a complex-valued Dirac delta-function. Now performing the standard procedure for the $P$-representation leads  to the conclusion that  the distribution $P(\alpha,\alpha^{\ast},t)$ cannot be a Dirac delta-function for any instant of time $t$, except for the initial instant of time $t=0$.\footnote{To this end see e.g. eq.(3.337) in \cite{breuer}.} The only exception to this finding is the initially zero state, $\vert\alpha_{\circ}\rangle\equiv\vert \alpha(t=0)=0\rangle$, for the case  $\hat H=\omega\hat a^{\dag}\hat a$,. However, this state returns the rhs of eq.(\ref{eqn:7}) to equal zero, thus being a stationary state for the process ($\hbar=1$):

\begin{equation}\label{eqn:18}
\mathcal{H}\vert\alpha_{\circ}\rangle=\left(\omega \hat a^{\dag}\hat a + \imath\gamma \langle\alpha_{\circ}\vert\hat a^{\dag}\vert \alpha_{\circ}\rangle \hat a
-{\imath\gamma\over 2} \langle\alpha_{\circ}\vert\hat a^{\dag}\hat a\vert \alpha_{\circ}\rangle -
{\imath\gamma\over 2} \hat a^{\dag}\hat a
\right)
\vert\alpha_{\circ}\rangle =0.
\end{equation}

\noindent Therefore, there is not even a single dynamically-evolving pure state satisfying eq.(\ref{eqn:7}) and eq.(\ref{eqn:9}) for the process on  $T=0$.

\noindent {\it The case $T>0$}. Then both terms in the master equation are nonzero ($N>0$) giving rise to the two Lindblad operators, $\hat a$ and $\hat a^{\dag}$. It is easy to show that the pure-dynamics condition (\ref{eqn:9}) now gives:

\begin{equation}\label{eqn:19}
\langle \hat a^{\dag} \hat a\rangle + {N\over2N+1} = \langle \hat a\rangle \langle \hat a^{\dag}\rangle.
\end{equation}

Introducing $\vert u\rangle =\hat a\vert v\rangle$, the Cauchy-Schwartz inequality, $\vert\langle u\vert v\rangle\vert^2\le \langle u\vert u\rangle\langle v\vert v \rangle$, implies:  $\langle \hat a^{\dag} \hat a\rangle \nless \langle \hat a\rangle \langle \hat a^{\dag}\rangle$.  Since $N\ge 0$, validity of eq.(\ref{eqn:19}) requires $N=0$. However, the condition $N=0$ is equivalent with the condition $T=0$, which contradicts the initial assumption of the bath's nonzero temperature. Therefore, for the finite temperature, there is not a single pure state that could fulfill the condition eq.(\ref{eqn:9}) for the PPSD of the damped harmonic oscillator.

\subsection{Comments}

Explicit use of the conditions eq.(\ref{eqn:7}) and eq.(\ref{eqn:9}) is performed in Appendix B for some other models of the quantum Markovian processes known in the literature. For all the considered models we obtain the same conclusion obtained in Sections II.A and II.C: there is not even  a single trajectory in the Hilbert state space of the system. Therefore we find the  {\it a priori} arguments against existence of trajectory in the Hilbert state space of an {\it open} system. Bearing in mind Section II.A we can conclude that, for the considered Markovian models, initially pure state turns into a mixed state and does not purify in the finite time intervals.
It is worth stressing albeit probably obvious: our findings may be expected to fail for the non-Markovian processes, for which some damping factors $\gamma_i$ in eq.(\ref{eqn:7}) may take negative values, at least for some instants of time.

\section{Discussion}

	We do not claim generality of our main finding presented in the previous section. That is, we do not claim nonexistence of the Hilbert-space trajectories for all Markovian processes, not even to mention non-Markovian dynamics. To this end, as a slight generalization for the Markovian processes with the Hermitian Lindblad operators, equation (\ref{eqn:9}), reads $\sum_i \gamma_i (\Delta\hat L_i)^2=0$ where $\Delta\hat L_i$ is the standard deviation of $\hat L_i$. Then the only possibility to fulfill the equality is for the common eigenstate (if such exists) of all $\hat L_i$s. However, as it can be easily shown, the common eigenstates are stationary states for the process, thus not producing any trajectory in the Hilbert state space of the system.

We also do not claim that the result obtained in Section II might present a complete answer to the question (Q). As emphasized in the Introduction section, extension of the analysis to the isolated $S+E$ system introduces a pure state for every single element of the ensemble of $S+E$ systems, and still, phenomenologically, seems to require non-unique dynamics for the subsystems—for both $S$ and $E$ systems. That is, we have to admit a lack of information regarding dynamics of the total $S+E$ system. This conclusion straightforwardly applies to the initial mixed state of the $S+E$ system.
	
By following the general logic repeatedly presented in the Introduction section, at least formally, one can consider the following possibility as an answer to the question (Q) within the unitary-only theory: Dynamics of every single $S$ system could be established for the $S$ system by assuming that every single $S$ system {\it is in the mixed state} $\hat\rho$, which is described by a master equation for the $S$ system.

This position is not entirely new. For instance, in Quantum information science, it is often assumed that single qubits can be described by mixed states \cite{Nilsen}, There are proposals to completely avoid the concept of statistical ensemble \cite{defner} in the framework allowing derivation of the Born's rule \cite{zurekBorn}. Nevertheless, those approaches are practically fully kinematical and hence not directly tackling the question (Q), which is the main subject of this paper. Therefore we proceed independently by emphasizing the following virtues of our proposal:

\begin{itemize}
  \item Individuality of every single element of the initial ensemble is preserved due to  the same mixed quantum state for every element of the ensemble in a later instant of time;
  \item The pure and mixed states are regarded on the equal footing as elements of the Banach space of state of the system;
  \item Dynamics of every single system $S$ is well defined and stated by the related master equation;
  \item Subsystem's states should be regarded ''proper'' but {\it not mixed} in the sense of the statistical-ensembles interpretation. This position is supported by: (i) equal treatment of the pure and mixed states (as stated above), (ii) the ensemble of the $S$ systems is homogeneous (all elements of the ensemble are in the same state), and (iii) the state carries all the possible information about the system that can be acquired by a quantum measurement;
  \item Quantum state of every single element of the $S+E$ system is pure and can be obtained by  ''purification'' \cite{Nilsen} of the $S$ system's state. The total system's ensemble is  homogeneous (every single element of the ensemble is in the same pure state) and can be determined by the proper quantum measurements on the ensemble of the $S+E$ systems;
  \item Nonuniqueness of the purification procedure is a consequence of nonunique decomposition of  mixed states generally and therefore does not pose any of the problem for our proposal.
\end{itemize}

Summarized, our argument can be stated as follows: The mixed states of an individual quantum subsystem may be formally equally treated as the pure states in all respects, including dynamics of the subsystem. The above third item provides a clear answer to the question (Q) posed in the Introduction section. Since individual open systems are described by mixed states, such states do not concern any ensemble behind the single systems and thereby the puzzle posed by ''improper mixtures'' dissolves.

Whether satisfied with our proposal or not, it is yet rather simple and consistent. Possibly the only problem with description of a quantum subsystem is the fact that  joining mixed states to an individual system directly tackles the problem of interpretation of probability on, mathematically speaking, the level of the single trial of an experiment \cite{khren, omnes}. In order more closely to present this problem, let us resort to the following illustration.

\begin{figure*}[!ht]
\centering
    \includegraphics[width=0.6\textwidth]{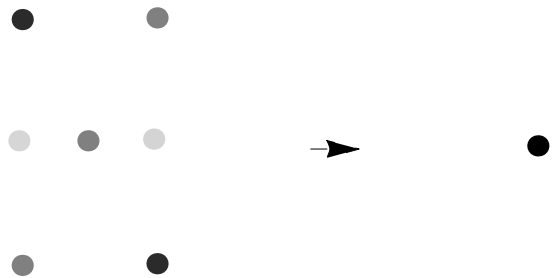}
\caption{Superimposed with the statistical weights of 1/6, the die sides give the “typical (average)” side presented on the left. Shadowing indicates the different statistical weights of the particular points in the pattern, which applies to every side of the die in the mixed state. The arrow indicates a “state reduction”, i.e. transition from the left (mixed state) to the right pattern (one of the a priori given sides (“pure states”) of the die), which (in the full color) appears with probability of 1/6.}\label{fig:1}
\end{figure*}

Let us consider the standard,  six-side die in a mixed state as illustrated by Fig.~{\ref{fig:1}}; statistical weights of $1/6$ for every side of the die is chosen. The left side of Fig.~{\ref{fig:1}} illustrates the “average” side of the standard die  in the mixed state that appears on {\it every} side of the die. Throwing the die is indicated by the arrow in Fig.~{\ref{fig:1}}.

In every single throw of the die, the initially mixed state (presented on the lhs of Fig.~{\ref{fig:1}}), {\it inescapably, and stochastically, turns out in to one}, and only one, single side out of the a priori known set of six sides of the die; the rhs of Fig.~{\ref{fig:1}} presents one possible outcome of throwing the die. This transition is indicated by the arrow in Figure 1, thus representing a kind of the (initially mixed) state reduction in to one ''pure state'' of the die. This situation is well known in the foundations of mathematical theory as the problem of interpreting probability  for the {\it single} trials of an “experiment”. Unpredictable appearance of a single side (a ''pure state'') of the die is typically described in the narrative as ''chance'', ''propensity'', ''emergence'' etc sometimes arguing for the non-deterministic physical world \cite{khren,gisin2, prigogine, auyang,gisin3}.
Now we can give the following  description of ''quantum measurement'':

\begin{itemize}
\item Initial tensor-product state, $\vert\phi\rangle_O\otimes\vert\chi\rangle_A$, of the object of measurement $O$ and the ''apparatus'' $A$ (which may be an open system, in which case the ''$A$'' regards the rest of the Universe for the $O$ system) completely describes ensembles of both $O$ and $A$ systems, as well as every single element of those ensembles. In the so-called pre-measurement stage, {\it interaction} between a single $O$ and a single $A$ system introduces entanglement for every such pair and dynamically leads to a mixed state $\hat\rho_O$, which does not admit any ensemble description. In the final stage of the measurement, a state reduction process, as illustrated by Fig.~{\ref{fig:1}}, stochastically transforms the mixed state $\hat\rho_S$ into one out of the a priori given set of possible pure quantum states. Those states should be regarded as stationary for the process and eigenstates of the measured observable.
\end{itemize}

\noindent Therefore, on the level of the object of measurement, the quantum measurement problem boils down to the problem of the proper understanding of the above described ''state reduction'' process. At this point, our proposal tackles the problem of interpretation of probability \cite{khren} (and the references therein). Needless to say, for the composite $O+A$ system, the measurement problem remains unanswered and can be traced back to the kinematical concept of ''improper mixtures'' (cf. the Introduction section). It is important to note: the statement (S), Section II, implies impossibility to ''unravel'' the subsystems quantum state in the sense that a mixed state cannot be {\it dynamically} regarded as a mixture of trajectories in the Hilbert state space. For completeness, we give the basic technical details that extend the considerations presented in Section III.

For a sufficiently smooth dynamics, imagined arbitrary decomposition (unraveling into the pure trajectories), $\hat\rho(t) = \sum_k p_k(t)$  $\vert\psi_k(t)\rangle$ $\langle\psi_k(t)\vert$, $\sum_i p_k(t)=1, \forall{t}$, in general,  implies:

\begin{equation}\label{eqn:20}
{d\hat\rho(t)\over dt} =
\sum_k \left(
{{dp_k(t)\over dt} \vert\psi_k(t)\rangle\langle\psi_k(t)\vert
+
p_k(t) {d\over dt} (\vert\psi_k(t)\rangle\langle\psi_k(t)\vert)}
\right).
\end{equation}

\noindent The knowledge of the (not necessarily orthogonal but normalized) $\vert\psi_k\rangle$s determines also the ''populations'' $p_k$ thus presenting eq.(\ref{eqn:20}) as a mathematically well defined problem.
Placing $p_k=1$ and $p_{k'}=0,\forall{k'\neq k}$ in eq.(\ref{eqn:20}) and substituting in eq.(\ref{eqn:6}) gives rise to the pure-state dynamics presented by equations (\ref{eqn:7}) and (\ref{eqn:8}).
Solutions to eq.(\ref{eqn:7})--if such exist--also determine the ''populations'' $p_k$ in eq.(\ref{eqn:20}).

Since existence of the $\rho_{ii}=\sum_k p_k \vert\langle i\vert\psi_k\rangle\vert^2$, $i=1,2,\dots, n-1$, $k=1,2,\dots, \mu$, while $\mu\le n-1$, is guaranteed by eq.(\ref{eqn:6}), and the states $\vert \psi_k\rangle$ are solutions to eq.(\ref{eqn:7}) (assuming that such exist), equation (\ref{eqn:20}) is well-defined algebraic equation for the unknown ''populations'' $p_k$ for every instant in time; by $n$ we denote dimension of the subsystem's Hilbert space.

Now, the knowledge of at least two pairs $(p_k(t),\vert\psi_k(t)\rangle)$, by definition,  ''unravels'' the imagined (although non-unique) ensemble
composition of the mixed state $\hat\rho$, which is subject of the dynamics eq.(\ref{eqn:6}). However, nonexistence of the trajectories $\vert\psi(t)\rangle$ presented by the statement (S), Section II, makes the equation (\ref{eqn:20}) {\it unphysical}: unraveling the improper mixtures is non-physical a task. This is implicit to the ERT method \cite{McCaul} for solving the Markovian master equations.

The concept of trajectory investigated in this paper regards the unitary-only dynamics and therefore has nothing in common with the so-called unraveling  into quantum-jump trajectories \cite{plenio} as well as with the CSL models of the objective state collapse \cite{bassi}. Also, the imagined trajectories within the de Broglie-Bohm theory \cite{durr} remain untouched by our considerations.

As a direct consequence of the statement (S), Section II, we recognize application of the (S) statement onto the idea of ''quantum history'' in the Hilbert state space of the system \cite{grifits}. Actually, a (fine-grained)\footnote{Extensions to the coarse-grained histories and for mixed initial state are straightforward.} history  is defined as a chain of projectors onto the Hilbert space, $\mathcal{C}=\hat P_{t_n}(t_n)\dots \hat P_{t_2}(t_2)\hat P_{t_1}(t_1), t_i>t_{i-1}$, and is in unique relation to a ''coarse grained'' trajectory in the Hilbert  space. That is, since every projector $\hat P_{t_i}(t_i)$ projects onto a pure  (albeit un-normalized) state, a dynamical chain for the history $\mathcal{C}$ is obvious:

\begin{equation}\label{eqn:C}
\vert\psi\rangle\to\vert\psi_{t_1}\rangle\to\vert\psi_{t_2}\rangle\to \dots\to\vert\psi_{t_n}\rangle,
\end{equation}

\noindent where $\vert\psi_{t_i}\rangle=\hat P_{t_i}(t_i)\vert\psi_{t_{i-1}}\rangle$, and $\vert\psi_{t_1}\rangle=\hat P_{t_1}(t_1)\vert\psi\rangle$ for the initial pure state $\vert\psi\rangle$. Needless to say, every state $\vert\psi_{t_i}\rangle$ in eq.(\ref{eqn:C}) is a point in a continuous trajectory in the Hilbert state space. Bearing the statement (S) in mind, we conclude that the models considered in Section II do not admit a definition of ''history'' for the considered open-systems Markovian models.

\section{Comparison with some alternative theories and interpretations of quantum theory}

Providing a systematic comparison of our proposal presented in Section III with the alternative theories and interpretations of quantum theory is far beyond the scope of this paper. Therefore, we will give some brief notions in this regard, starting from a brief exposition of the problem illustrated by Fig.~{\ref{fig:1}}.

\subsection{The concept of probability of a single physical system}

The process illustrated by Fig.~{\ref{fig:1}} is an old issue, probably as old as the attempts of describing and introducing the concept of probability \cite{khren} (and the references therein). The state transition presented in Fig.~{\ref{fig:1}} is typically said to be random, although the very concept of randomness carries a load of different meanings, mathematically as well as physically. In any way, physically, it seems to require rethinking the very basic physical laws, both classical and quantum ones. For instance, if the transition in Fig.~{\ref{fig:1}} is fully random (understood as unpredictable), then it is hard to even think of any underlying, mathematically descriptive mechanism behind the state transition. In this spirit, we have repeatedly emphasized that an alternative to our proposal may be the absence of  dynamical law for the individual systems subject of the considered Markovian processes.

While the pure randomness is in its essence lawless and therefore speculative, there is a hope for detecting dynamical laws (dynamical maps) for individual quantum subsystems, at least in principle. Before we learn about such law(s), we suggest to resort to our proposal as it reduces to the classical probability theory that seems to admit thinking in the terms of the ''proper'' \cite{Despa} statistical ensembles: an instantaneous mixed state $\hat\rho$ may be {\it treated} as a statistical mixture (ensemble) of pure states {\it just before} the mixed-state reduction, Fig.~{\ref{fig:1}}. While the choice of the final pure state may seem mysterious or even miraculous (if only due to nonunique ensemble decomposition of $\hat\rho$), this choice may intuitively seem less radical that the old Copenhagen ''quantum state collapse'' or the ''world branching'' in the Everettian interpretation(s) \cite{wallace}.

As long as we lack a general mathematical basis for the individual-system probability, it seems that the problem cannot be separated from certain interpretational contexts. As one such possibility appears the possibility of the so-called top-down causation, which deals with the composite system, $S+E$ (or $O+A$) \cite{khren,prigogine,auyang,gisin3}. This view fits with our proposal, which regards the quantum open systems (subject of the considered Markov processes): in principle, the mixed state reduction, Fig.~{\ref{fig:1}}, may bear on some kind of the top-down causation. Whether or not this can be designed so as to preserve unitarity of the total system's dynamics remains an open question yet. Without further ado, we finish this general consideration of our proposal presented in Section III.

\subsection{Some interpretations of the unitary-only theory}

Our considerations are purely on the level of the standard quantum formalism and therefore does not seem to contribute or to have much to learn from the de Broglie-Bohm hidden-variables interpretation \cite{durr} of quantum theory.

On the other hand, non-viability of the concept of ''history'', see equation (\ref{eqn:C}), is particularly interesting since the models analysed in Section I are  models of quantum decoherence (measurement), which is often used in modern Everettian Many Worlds Interpretation as a basis for quasi-classicality and therefore consistency of the interpretation \cite{wallace}. Bearing in mind the statement (S) of Section II, whatever might be the subsystem's dynamics, it might be better not to be Markovian. That is, it would be very interesting to find a counterexample to the statement (S) thus possibly allowing for the concept of ''history'' and particularly of the ''consistent history'' \cite{grifits}, which is essential for the Many Worlds Interpretation (emergent or non-emergent Worlds and hence of emergent or non-emergent subsystems) \cite{wallace}. Finally, we find some similarity of the reduction process, see Fig.~{\ref{fig:1}}, with the choice for branching of an instantaneous single Everettian quantum world--some ''decision'' allowing the classical thinking should be made in both cases. The similarity is of course formal, since there is no need for ''rational agent'' \cite{dojc} within our proposal exposed in Section III.

\subsection{Some objective collapse theories and interpretations}

The stochastic processes behind the objective collapse (reduction) of state \cite{kastner,Bassi} are of the same general kind as the state reduction in Fig.~{\ref{fig:1}}. The basic distinction lies in the fact that the assumed stochastic processes make the overall dynamics of the $S+E$ system non-unitary. Such an assumption is not a part of our proposal that simultaneously exhibits the possible limits of the unitary-only quantum theory.  In any case the output state is a ''proper mixture'' \cite{Despa} allowing for the standard, ignorance-based statistical ensemble description.
It is a nontrivial observation, that our proposal does not call for any stochastic {\it process} repeatedly occurring in some instants of time. Rather, our proposal assumes the occurrence of the state reduction only in a single time instant thus not necessarily presenting a stochastic {\it process}; the state reduction in our proposal may as well be a quick, non-instantaneous process.
Regarding our proposal it remains open if the state reduction can be regarded a part of the unitary-only theory, at least on the level of interpretation. To this end, as already emphasized above, we may need some hints from mathematical considerations of the concept of probability.

\subsection{Taking the individual system's mixed states seriously}

In quantum information science (QIS), it is customary to regard single qubits in mixed state \cite{Nilsen}. This may be basically due to the strong operational view of quantum theory of, possibly, majority of the quantum information community. In any case, operationally, the measurement probabilities are assumed describable in the frequentist manner with the well defined initial state preparation. Therefore our proposal goes hand in hand with this QIS wisdom while suffering from the same problem of the state reduction presented in Fig.~{\ref{fig:1}}. Neither QIS nor our standpoint are based on any interpretation.

A much more radical adoption of mixed states for individual systems can be found within a proposal of totally abandoning the concept of statistical ensemble, probability, randomness and all that \cite{defner}. Nevertheless, as compared to our proposal, this kinematical approach to statistical physics foundations seems to be incomplete. Particularly, it is not clear if dynamical transition from an initial to an actual (entangled) state is irrelevant. If relevant, then it may turn out that the proposal  requires some additional ingredients, in spirit, possibly closer to our proposal.

\subsection{A Local Time perspective}

Our considerations started from the trivial observation, cf. Introduction section, that the problem imposed by improper mixtures begins as soon as the interaction of an open system $S$ starts to unfold with some external system $E$. This notion qualitatively fully complies with the concept of ''sufficiently strong'' interaction that is essential for the so-called Local Time Scheme (LTS) \cite{JJD1,JJD2,Hitoshi}, which assumes  unitary dynamics for every single $S+E$ system. The very concept of local time \cite{JJD1,JJD2,Hitoshi} introduces stochasticity on the level of single composite $S+E$ system, without affecting unitary dynamics for the total system while providing, as expected, mixed state for the $S$ system. While pending for the further analysis of the LTS, we can recognize sustainability of our proposal (Section III) within the basics of the Local Time Scheme.

\section{Conclusion}

Dynamics consisting exclusively of the transitions from a pure into another pure quantum state is forbidden for certain models of Markovian open system models, which include some important models of quantum decoherence and measurement.  Assigning mixed quantum states to individual systems (to individual elements of a statistical ensemble)  dissolves the problem of ''improper mixture'', establishes the open systems dynamics and presents the quantum measurement problem in a new guise. If unsatisfied with this proposal, we may face the possibility that the very concept of dynamics of open quantum systems cannot be properly introduced.

\begin{acknowledgments} The present work was supported by The Ministry of Education, Science and Technological Development of the Republic of Serbia (451-03-68/2020-14/200122)
and in part for MD by the ICTP–-SEENET-MTP project NT-03 Cosmology--Classical and Quantum Challenges. We are indebted to Aurelian Isar for the discussion on the subject.
\end{acknowledgments}

\appendix

\section{The PPSD  equation}

For completeness, we outline the main steps leading to eqs.(\ref{eqn:7}) and (\ref{eqn:9}) as it can be found in \cite{sandu}.

Assume the pure state $\hat\rho(t)=\vert\varphi(t)\rangle\langle\varphi(t)\vert, \forall{t}$. Introduce arbitrary pure state (a time-independent vector) $\vert\theta\rangle$. Then $\hat\rho\vert\theta\rangle=\langle\varphi\vert\theta\rangle \vert\varphi\rangle$ so it is straightforward to obtain:

\begin{equation}
{d\over dt}\left(\hat\rho\vert\theta\rangle\right) = \mathcal{L}[\hat\rho] \hat\rho \vert\theta\rangle + \hat\rho\mathcal{L}[\hat\rho]  \vert\theta\rangle.
\end{equation}

Now the use of $\hat\rho \ast \hat\rho = (tr \ast\hat\rho) \hat\rho$ and the, possibly time dependent, Liouvillian $\mathcal{L}$ from eq.(\ref{eqn:6}), straightforwardly lead to the Schr\" odinger-like equation \cite{sandu}:

\begin{equation}
{d\vert\varphi\rangle\over dt} = -{\imath\over\hbar}\mathcal{H} \vert\varphi\rangle,
\end{equation}

\noindent which is equivalent with eq.(\ref{eqn:7}) and eq.(\ref{eqn:8}).

\section{Models of Markovian processes}

Following the considerations in Section II.C, we analyse some standard Markovian master equations. Our analysis formally includes the master equations that regard the isolated quantum systems, cf. the models 8 and 9 below. The findings extend the findings in Sections II.A and II.C.

\subsection{A three-level atom}

A three-level atom with the (non-degenerate) energies $E_1 < E_2 < E_3$ is endowed by the dipole transitions which exclude the transitions between the two lower levels. Defining
the operators $\hat \sigma_{ij}\equiv\vert i\rangle\langle j\vert, i\neq j=1,2,3$,
the master equation (in the interaction picture) follows from the quantum-optical master equation \cite{breuer,Gardiner}:

\begin{equation}
\nonumber
\dot{\hat\rho} = \gamma_1 (N_1+1) \left(\hat\sigma_{13}\hat\rho\hat\sigma_{31}-{1\over 2}\{\hat\sigma_{31}\hat\sigma_{13},\hat\rho\}\right) +
 \gamma_1 N_1 \left(\hat\sigma_{31}\hat\rho\hat\sigma_{13}-{1\over 2}\{\hat\sigma_{13}\hat\sigma_{31},\hat\rho\}\right) +
 \end{equation}
\begin{equation}
\gamma_2 (N_2+1) \left(\hat\sigma_{23}\hat\rho\hat\sigma_{32}-{1\over 2}\{\hat\sigma_{32}\hat\sigma_{23},\hat\rho\}\right) +
   \gamma_2 N_2 \left(\hat\sigma_{32}\hat\rho\hat\sigma_{23}-{1\over 2}\{\hat\sigma_{23}\hat\sigma_{32},\hat\rho\}\right).
\end{equation}

\noindent In eq.(B.1): $N_i\equiv N(\omega_i)=(e^{\omega_i/k_BT}-1)^{-1}$, with the transition frequencies $\omega_i, i=1,2$; we assume $\hbar=1$.

Therefore the four Lindblad operators (for the finite temperature): $\hat\sigma_{13}, \hat\sigma_{31}$, $\hat\sigma_{23}, \hat\sigma_{32}$; generalization to the degenerate case is straightforward.
Then the PPSD condition eq.(\ref{eqn:9}) of the main text gives the equality:

\begin{equation}
\gamma_1N_1(p_1+p_3-2p_1p_3) + \gamma_1(p_3-p_1p_3) + \gamma_2N_2(p_2+p_3-2p_2p_3) + \gamma_2(p_3-p_2p_3)=0,
\end{equation}

\noindent where $p_i\equiv\vert\langle i\vert\psi\rangle\vert^2$.

\noindent {\it The case $T=0$.} For $T=0$,  $N_1 = 0 = N_2$  and eq.(B.2) requires $p_3=0$, with arbitrary $p_i,i=1,2$. That is, every pure state $\vert\psi\rangle = c_1\vert 1\rangle + c_2\vert 2 \rangle$ is allowed.
However, since $\hat\sigma_{i3}\vert\psi\rangle=0, \forall{i=1,2,3}$, placing $N_1=0=N_2$ into eq.(B.1) reveals that every pure state $\vert\psi\rangle$ (i.e. for arbitrary $c_i, i=1,2$) is a stationary state.
That is, such states do not evolve in time and hence there is not even a single pure state satisfying eq.(\ref{eqn:9}) of the body text.

\begin{figure*}[!ht]
\centering
    \includegraphics[width=0.6\textwidth]{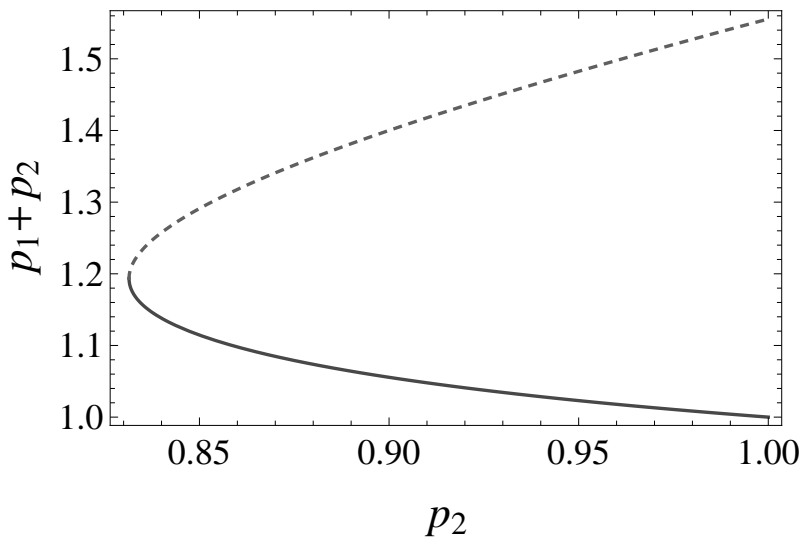}
\caption{The plot of $p_1+p_2$ for the  nonzero $p_3$. The dashed line is for the one, while the solid one is for the other solution for $p_1$ following from eq.(B.3). It is obvious that $p_1+p_2>1$ for every choice of $p_2>0.83$.}\label{fig:2}
\end{figure*}

\noindent {\it The case $T>0$.} Then $N_1\neq 0 \neq N_2$ and the pure-state condition reads:

\begin{equation*}
\gamma_1(N_1+1)(p_3-p_1p_3) + \gamma_1N_1(p_1-p_1p_3)+\gamma_2(N_2+1)(p_3-p_2p_3)+\gamma_2N_2(p_2-p_2p_3)=0, \tag{B.3}
\end{equation*}

\noindent with the constraint $p_1+p_2+p_3=1$. Without loss of generality, choose (like for the models of the atomic dark states and induced transparency \cite{Gardiner,Scully}) $\gamma_1=1=100\gamma_2$ and $N_1=0.4=1000N_2$.
By inspection it can be seen that eq.(B.3) does not return any relevant solutions. For example, place $p_3=1-p_1-p_2$ and solve eq.(B.3) for $p_1$. For $p_3=0$, eq.(B.3) implies a negative value for either $p_1$ or $p_2$. For nonzero $p_3$, Fig.~{\ref{fig:2}} exhibits that the sum $p_1+p_2>1$; the minimum value for $p_2$ that returns a real value for $p_1$ is larger than $0.83$.

Analogous elimination of the other terms (either $p_1$ or $p_2$) leads to the similar conclusions. Therefore it is found, that even a single pure state
cannot satisfy eq.(\ref{eqn:9}).

\subsection{A multimode system}

Consider a linear system of $n$ harmonic oscillators or more generally modes in contact with a thermal bath on the temperature $T$. In any case this system can be transformed into a set of mutually uncoupled normal coordinates,
 i.e. of uncoupled modes, here presented by the commuting Bose annihilation operators $\hat a_i, i=1,2,...,n$; $[\hat a_i,\hat a_j^{\dag}]=\delta_{ij}$. Then a generalization of eq.(\ref{eqn:16}) of the main text is straightforward:

\begin{equation}
\dot{\hat\rho} = \sum_{i=1}^n \gamma_i \left(
(N(\omega_i)+1)\left(\hat a_i\hat\rho \hat a_i^{\dag}-{1\over 2}\{\hat a_i^{\dag}\hat a_i,\hat\rho\}\right) + N(\omega_i) \left(\hat a_i^{\dag}\hat\rho \hat a_i-{1\over 2}\{\hat a_i\hat a_i^{\dag},\hat\rho\}\right)
\right). \tag{B.4}
\end{equation}

The Lindblad operators are all the Bose operators, $\hat a_i$ and $\hat a_i^{\dag}$. Then eq.(\ref{eqn:9}) takes the form:

\begin{equation}
\sum_i \gamma_i \left( (N(\omega_i)+1)
\left(\langle \hat a_i^{\dag} \hat a_i \rangle - \langle \hat a_i\rangle\langle \hat a_i^{\dag}\rangle\right) +
N(\omega_i) \left(\langle \hat a_i \hat a_i^{\dag}\rangle - \langle \hat a_i^{\dag}\rangle \langle \hat a_i\rangle \right)
\right)=0. \tag{B.5}
\end{equation}

Due to the Cauchy-Schwarz inequality (cf. the main text), as well as to $\gamma_i, N(\omega_i) \ge 0, \forall{i}$, all the terms in eq.(B.5) are non-negative. Therefore the only possibility to fulfill the equality in (B.5) is already recognized in Section II.C of the main text:

\begin{equation}
N(\omega_i) = 0, \quad \langle \hat a_i^{\dag}\hat a_i \rangle - \langle \hat a_i\rangle\langle \hat a_i^{\dag}\rangle=0, \tag{B.6}
\end{equation}

\noindent for every mode $i$. Mutual independence of the modes implies the conclusion drawn in Section II.C for every mode $i$ as well as for the solution, $\otimes_i\vert\alpha_i\rangle$, of eq.(B.5) (where $\alpha_i$
states for the $i$th ''coherent state'' for the $i$th mode).

\subsection {The phase damped harmonic oscillator}

This model \cite{Gardiner} is comprised of a single harmonic oscillator in contact with a thermal bath such that the bosonic number operator, $\hat N=\hat a^{\dag}\hat a$, is coupled with the bath's variable(s).
The effective master equation (in the Schr\" odinger picture) is ($\hbar=1$):

\begin{equation}
\dot{ \hat\rho }= -\imath \omega_{\circ} [\hat N,\hat\rho] +\gamma \left(\hat N\hat\rho \hat N - {1\over 2}\{{\hat N}^2,\hat\rho\}\right) \tag{B.7}
\end{equation}

The only Lindblad operator, i.e. the number operator, $\hat N$ commutes with the system Hamiltonian and hence the PPSD condition:

\begin{equation}
(\Delta \hat N)^2=\langle\hat N^2\rangle - \langle\hat N\rangle^2=0, \tag{B.8}
\end{equation}

\noindent determines its eigenstates $\vert n\rangle$, $\hat N\vert n\rangle = n\vert n\rangle$, as the solutions. However, those states are the exact ''pointer basis'' states \cite{Giulini} that do not evolve in time. Thus there is not a single pure state
whose dynamics could describe the decoherence dynamics eq.(B.7). This conclusion applies to all the similar Markovian decoherence-models for a qubit or a continuous-variable (CV) systems \cite{breuer,Giulini,Slosi}.

\subsection {The one-qubit generalized depolarizing channel}

The so-called generalized one-qubit depolarizing channel is modelled by the following master equation (in the interaction picture) \cite{BJP17}:

\begin{equation}
\dot {\hat\rho} = \sum_{i=x}^z \gamma_i (\hat\sigma_i\hat\rho \hat\sigma_i - \hat\rho), \tag{B.9}
\end{equation}

\noindent where appear the standard Pauli operators $\hat\sigma_i, i = x,y,z$.

Therefore there are three Lindblad operators, $\hat L_i=\hat\sigma_i$, which give rise to the PPSD condition:

\begin{equation}
\sum_i \gamma_i (1-\langle \hat\sigma_i\rangle^2) = 0.  \tag{B.10}
\end{equation}

It is obvious that eq.(B.10) implies $\langle\hat\sigma_i\rangle=1,\forall{i}$, which cannot be fulfilled for any pure state.
This result may be expected due to the fact that this is a unital channel, i.e., dynamics preserving the fully mixed state, $\hat I/2$, where $\hat I$ is the identity operator .
This is a situation also for all the unital maps for which the Lindblad operators do not have even a single common eigenstate \cite{breuer}.

\subsection {Decay of a two-level atom into a squeezed field vacuum}

This is another standard model in quantum optics for a two-level atom that is described by the following master equation \cite{breuer,Gardiner}:

\begin{equation}
\dot {\hat\rho} = \gamma_{\circ} \left(\hat
C\hat\rho\hat C^{\dag} - {1\over 2} \{\hat C^{\dag}\hat C,\hat\rho\}
\right), \tag{B.11}
\end{equation}

\noindent where $\hat C = \cosh(r) \hat\sigma_- + e^{\imath \theta}\sinh(r) \hat\sigma_+$, and the environmental squeeze parameters $r$ and $\theta$, while $\hat\sigma_-=\vert g\rangle\langle e\vert$ for the excited ($e$) and the ground ($g$)
atomic states.

Since there is only one Lindblad operator, $\hat L=\hat C$, the PPSD condition reads:
\begin{equation}
\cosh^2(r) p_g + \sinh^2(r)(1-p_g) - p_g(1-p_g)(\cosh(2r)+\sinh(2r)\cos(\omega)) = 0; \tag{B.12}
\end{equation}

\noindent in eq.(B.12): $p_g=\vert \langle g\vert\psi\rangle\vert^2=1-p_e$, $\omega = \theta+2\delta$, with the arbitrary phase $\delta$.

Solutions to eq.(B.12) are readily obtained: $(1 + e^{\imath \omega} \coth(r))^{-1}$ and $1-(1+e^{\imath\omega}\tanh(r))^{-1}$. The complex term can be eliminated
for either $\omega=0$ or $\omega=\pi$. For the latter the negative  values for $p_g$ are obtained. Therefore the only possibility is $\omega=0$ (i.e. $\delta=-\theta/2$) when the two solutions become equal. Hence
the unique solution  $p_g = (1 + \coth(r))^{-1}$, which is independent of $\theta$. Thus, for the  environmental  state with the fixed $r$ and $\theta$ parameters, the {\it unique} state of the two-level
system is found to read:

\begin{equation}
\vert\psi\rangle = (1 + \coth(r))^{-1/2} \vert g\rangle + e^{-\imath\theta/2} (1-(1 + \coth(r))^{-1})^{1/2} \vert e\rangle, \tag{B.13}
\end{equation}

\noindent which is the eigenstate of the operator $\hat C$ with the eigenvalue $e^{\imath\theta/2}\sqrt{\sinh(2r)/2}$, as it can be easily checked.

Certainly, the single pure state eq.(B.13) cannot provide even a single trajectory in the Hilbert space of the system. For completeness, below we give a solution to the master equation (B.11).
Going to the matrix representation of the dissipator, $\mathcal{D}$, in eq.(B.11) and then to the map $e^{\mathcal{D}t}$. The state given in the general form $\hat\rho(t)=(\hat I + \sum_i n_i(t) \hat\sigma_i)/2$,
where $n_i(t)=tr(\hat\rho(t)\hat\sigma_i)$, for the initially pure state satisfies: $n_x(0)=2\sqrt{p_e(1-p_e)}\cos(\delta), n_y(0)=-2\sqrt{p_e(1-p_e)}\sin(\delta)$ and $n_z(0)=2p_e-1$, while $n_x^2(0)+n_y^2(0)+n_z^2(0)=1$, for arbitrary phase $\delta$ and $p_e=\vert\langle e\vert\psi\rangle\vert^2$. Then the solutions for the Bloch vector components can be found:

\begin{widetext}
\begin{eqnarray}
&\nonumber&
n_x(t) = e^{-{\gamma t\over 2}e^{2r}} \left((1-\cos(\delta)+e^{\gamma t\sinh(2r)}(1+\cos(\delta)))n_x(0)+\sin(\delta)(1-e^{\gamma t\sinh(2r)})n_y(0)  \right)
\\&&\nonumber
n_y(t) = e^{-{\gamma t\over 2}e^{2r}} \left(\sin(\delta)(1-e^{\gamma t\sinh(2r)})n_x(0)+(1+\cos(\delta)+e^{\gamma t\sinh(2r)}(1-\cos(\delta)))n_y(0)  \right)
\\&&\nonumber
n_z(t)= e^{-2\gamma t\cosh^2(r)}n_z(0).
\end{eqnarray}
\end{widetext}

\begin{figure*}[!ht]
\centering
    \includegraphics[width=0.6\textwidth]{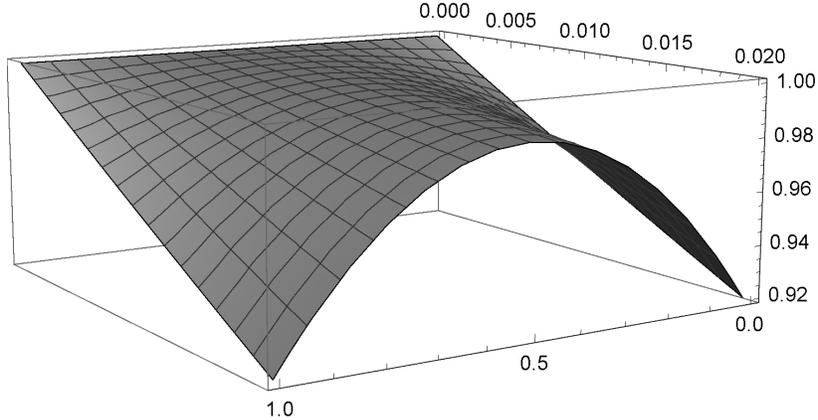}
\caption{The plot of the sum, $P=n_1^2+n_2^2+n_3^2$, for the choice of the parameters: $\gamma_{\circ}=1,r=0.2,\theta=\pi,\delta=-\pi/2$.}\label{fig:3}
\end{figure*}

In Fig.~{\ref{fig:3}} we graphically present the sum $P(t)=n_x^2(t)+n_y^2(t)+n_z^2(t)$, which is assumed to take the initial value $P(0)=1$. Without loss of generality, we choose: $\gamma_{\circ}=1, r=0.2,\theta=\pi$ and $\delta=-\pi/4$ (the choice of $\theta=-2\delta$ complies with the state eq.(B.13)).

\noindent Obviously, the condition of purity of state, $P=n_1^2+n_2^2+n_3^2=1$, is fulfilled for the initial instant of time $t=0$, and, from Fig.~{\ref{fig:3}}, for $t>0$ {\it only} for $r=0$, which is the case $\gamma_1=0$ investigated in Section II.C. Therefore there is not even a single trajectory in the Hilbert state space for the process.

\subsection {A nonadiabatic Markovian model}

For an externally driven damped harmonic oscillator, the following Markovian master equation applies (in the interaction picture) \cite{Dann}:

\begin{equation}
\dot{\hat\rho} = \vert\xi(t)\vert^2 \gamma(t) \left(
\hat F_+\hat\rho \hat F_- - {1\over 2}\{\hat F_- \hat F_+,\hat\rho\} + e^{-\hbar\alpha(t)/k_BT}
(\hat F_-\hat\rho \hat F_+ - {1\over 2}\{\hat F_+\hat F_-,\hat\rho\})
\right), \tag{B.14}
\end{equation}

\noindent where $\hat F_+= A \hat x + B \hat p = \hat F_-^{\dag}$ and $A=(1+\imath\mu/\kappa)/2, B=\imath/m\omega(0)\kappa$, with all the positive parameters, $m,\omega(0),\kappa >0$ as well as $\gamma(t), \alpha(t)\ge 0, \forall{t}$.

From (B.14), the two Lindblad operators are found: $\hat L_1 = \hat F_+$ and $\hat L_2=e^{-\hbar\alpha(t)/2k_BT}\hat F_-$. Then applying eq.(\ref{eqn:9}) from the main text gives:

\begin{equation}
\left(
\langle \hat F_-\hat F_+\rangle - \langle \hat F_-\rangle \langle \hat F_+\rangle
\right)(1+ e^{-\hbar\alpha(t)/k_BT}) +  e^{-\hbar\alpha(t)/k_BT}  \langle [F_+,F_-]\rangle=0.  \tag{B.15}
\end{equation}

With the use of the commutator, $[\hat F_+,\hat F_-]=\hbar/m\omega(0)\kappa$, eq.(B.15) becomes a sum of the non-negative terms:

\begin{equation}
\left(
\langle \hat F_-\hat F_+\rangle - \langle \hat F_-\rangle \langle \hat F_+\rangle
\right)(1+ e^{-\hbar\alpha(t)/k_BT}) +   {\hbar  e^{-\hbar\alpha(t)/k_BT}\over m\omega(0)\kappa}=0.  \tag{B.16}
\end{equation}

Since the second term on the rhs of (B.16) cannot equal zero, we conclude that there does not exist even a single pure state for the master equation (B.14) that could satisfy eq.(\ref{eqn:9}).

\subsection {The Walls-Collet-Milburn model}

Consider the dynamical model of quantum measurement performed on a two-mode system (a system $S$ with the mode $\hat a$ and the ''meter'' system $M$ with the mode $\hat b$) \cite{Walls}, such that only one of the coupled modes is in contact with the thermal bath of modes $\hat c_i$ on some temperature $T$.
Assuming that  interaction of the object of  measurement with meter system  is given by a four-wave-mixing interaction:

\begin{equation}
\hat H_{SM} = - {\imath\over 2} \hat a^{\dag} \hat a \otimes (\epsilon^{\ast}\hat b - \epsilon\hat b^{\dag}) \tag{B.17}
\end{equation}

\noindent and that the interaction of the meter with the interaction is bilinear:

\begin{equation}
\hat H_{ME}= \hat b\otimes \hat C^{\dag} + \hat b^{\dag} \otimes \hat C, \tag{B.18}
\end{equation}

\noindent where $\hat C^{\dag} = \sum_j\kappa_j \hat c_j$, it can be shown that, for longer times, the master equation for the object of measurement (in the interaction picture) is of the Markovian form:

\begin{equation}
\dot{\hat{\rho}}_S = {\vert\epsilon\vert^2\over\gamma}\left(\hat N\hat\rho\hat N-{1\over 2} \{\hat N,\hat\rho_S\}\right) \tag{B.19}
\end{equation}

\noindent where $\hat N \equiv \hat a^{\dag}\hat a$ and the real parameter $\gamma>0$.

In eq.(B.19) appears only one Lindblad operator, $\hat L\equiv\hat N$, which placed in eq.(\ref{eqn:9}) gives rise to the following condition for the pure-state dynamics of the object of measurement:

\begin{equation}
(\Delta\hat N)^2 \equiv \langle\hat N^2\rangle - \langle\hat N\rangle^2 = 0. \tag{B.20}
\end{equation}

Of course, every eigenstate, but no linear combination of the eigenstates, of the number operator $\hat N$ satisfies the condition eq.(B.20). However, as it can be easily seen,
those states also represent the stationary states for the process, thus not being subject of any change in time.

\subsection {Quantum mechanics with spontaneous localization model}

Consider the Ghirardi-Rimini-Weber master equation \cite{Bassi}:

\begin{equation}
\dot{\hat{\rho}} = \lambda \left(
\sqrt{{\alpha\over\pi}} \int ds e^{-\alpha(\hat x-s)/2}\hat\rho e^{-\alpha(\hat x-s)/2} - \hat\rho
\right) \tag{B.21}
\end{equation}

\noindent It is worth emphasizing that equation (B.21) is assumed to describe dynamics of a {\it closed}, not of an open, system.

There  is a Lindblad operator, $\hat L_s\equiv e^{-\alpha(\hat x-s)/2}$, for every value of the continuous real parameter $s$. Then the condition eq.(\ref{eqn:9}) obviously implies:

\begin{equation}
\Delta (e^{-\alpha(\hat x-s)^2/2}) = 0, \forall{s} \tag{B.22}
\end{equation}

\noindent for the standard deviation of the Lindblad operators.

Needless to say, there is not even a single pure state (in the Hilbert space) that could fulfill eq.(B.22) and hence nonextistence of even a single trajectory in the Hilbert state space of the system.

\subsection {Continuous spontaneous localization model}

The master equation of interest that is assumed to describe dynamics of a {\it closed} system reads \cite{Bassi}:

\begin{equation}
\dot{\hat{\rho}} = \lambda \sum_{i,k} \left(
\hat N_i^{(k)} \hat\rho \hat N_i^{(k)} - {1\over 2} \{\hat N_i^{(k)2},\hat\rho\}
\right), \tag{B.23}
\end{equation}

\noindent with the Hermitian Lindblad operators $\hat N_i^{(k)}$, which represent appropriate bosonic number operators for the model. Then eq.(\ref{eqn:9}) takes the form:

\begin{equation}
\sum_{i,k} (\Delta\hat N_i^{(k)})^2 = 0, \tag{B.24}
\end{equation}

\noindent where appear the standard deviations for $\hat N_i^{(k)}$s. Certainly, eq.(B.24) can be only identically fulfilled, i.e. only if $\Delta \hat N_i^{(k)}=0, \forall{i,k}$. Therefore,
the only pure states allowing in principle existence of trajectories are the common eigenstates for all $\hat N_i^{(k)}$s. However, it can be easily shown than the eigenstates for those operators, denoted $\vert n_1^{(k)} n_2^{(k)} n_3^{(k)} ... \rangle$, when placed into eq.(B.23) return the rhs of eq.(B.23) to be equal to zero. That is, such states are stationary states that do not evolve in time.


\begin{thebibliography}{9}
\bibitem{foot}
Here we stick to the terminology used e.g. in \cite{kastner} and indicating that an isolated quantum system is subject, only and exclusively, of the unitary Schr\" odinger dynamics.
\end{thebibliography}
\end{document}